\documentclass[12pt]{article}
\usepackage[authoryear,sort]{natbib}
 \usepackage[margin=1in]{geometry}
\usepackage{amsmath}
\usepackage[affil-it]{authblk}
\usepackage{blindtext}
\usepackage{verbatim}

\usepackage{tikz}

\usepackage{hyperref}
 \hypersetup{
     colorlinks=true,
     linkcolor=blue,
     filecolor=blue,
     citecolor = blue,      
     urlcolor=cyan,
     }
\usepackage{amsthm}
\usepackage{amssymb}
\usepackage{xcolor}
\usepackage[english]{babel}

\newtheorem{thm}{Theorem}
\newtheorem{prop}{Proposition}
\newtheorem{definition}{Definition}
\newtheorem{lemma}{Lemma}
\newtheorem{example}{Example}[section] 
\title{Observational Learning with Competitive Prices\footnote{I would like to thank my advisor, Prof. Navin Kartik, for all his help and guidance. I also appreciate useful comments and suggestions from Prof. Steven Ho.}}

\author{Zikai Xu
	\thanks{Department of Economics, Columbia University. E-mail: xu.zikai@columbia.edu}
		}

\date{\today}

\begin{document}
	
	\maketitle
	\begin{abstract}
		A market gets learning when rational agents in the market learn the value of underlying asset correctly and asymptotically. This paper studies observational learning in a market with competitive prices, and explores conditions for market learning. Comparing a market with public signals and a market with private signals in a sequential trading model, we find that pairwise informativeness (PI) is a sufficient and necessary learning condition for a market with public signals; and Avery and Zemsky Condition (AZC) is a sufficient and necessary learning condition for a market with private signals. Moreover, when the number of states is 2 or 3, PI and AZC are equivalent. And when there are more than 3 states, PI and Monotonic Likelihood Ratio Property (MLRP) together imply market learning in the private signal case.
	\end{abstract}

\newpage
\tableofcontents
\newpage
\section{Introduction}
This paper considers observational learning processes where prices exist. An typical example is that in financial markets, agents observe history trading information and then make their own decisions. The question we ask is, when can agents be arbitrarily certain about the true value of the asset eventually? The same question is asked by \cite{avery1998multidimensional}, who have shown that under an assumption (which is named Avery and Zemsky Condition by this paper), the market with private signals will get learning. And this paper will discuss the necessity of AZC for market learning. The problem of AZC is that it involves not only signal structure, but also agents' beliefs, which makes it hard to test. This paper aims to characterize the signal structures that satisfies AZC without involving beliefs. The main results of this paper shows that pairwise informativeness is equivalent to AZC when no more than 3 states. And MLRP plus PI is sufficient for AZC. Here both PI and MLRP are conditions solely focusing on signal structures.

The learning model of financial markets initiated by \cite{glosten1985bid}, then developed by \cite{avery1998multidimensional} and \cite{park2011herding} is the main objective of this paper. Market makers dynamically set bid ask spread to earn zero expected profit because of competitive market. Informative agents trade if and only if under the belief that they can make strictly positive expected profit. Both informative agents and market makers are Bayesian decision makers sharing a common prior $\mu_0$. There are also noise agents who make decisions uniformly, so we do not care the belief dynamics of noise agents. At each date, market makers post bid price and ask price, then an agent makes a trading decision. To study the impact of signal structures on observational learning with competitive prices, this paper splits the traditional market learning model into a private signal case and a public signal case. In the private signal case, each date an informative agent comes with a known probability (otherwise a noise trader comes), and observes a private signal which is i.i.d conditional on true state. The market makers and the informative agent share the public beliefs before the private signal occurs. Upon a private signal occurs, the informative agent updates the public belief into her private posterior by Bayes rule.\footnote{The public belief $\mu_t$ at date $t$ is the belief of informative agents and market makers at the beginning of period $t$. It is the posterior belief of a Bayesian decision maker whose prior is $\mu_0$ and observing history information by date $t$.} Everyone else can only infer the interval where the private signal could be by observing her action. While in the public signal case, signals observed by informative agents perfectly "leak" out, that is, each private signal is open to all subsequent agents. The case when private information is partially revealed to the public is left for future research. Unlike previous research in market learning mainly studying short term herding behavior, this paper focuses on the asymptotic learning outcomes.

The main results of this paper can be presented in the figure below:
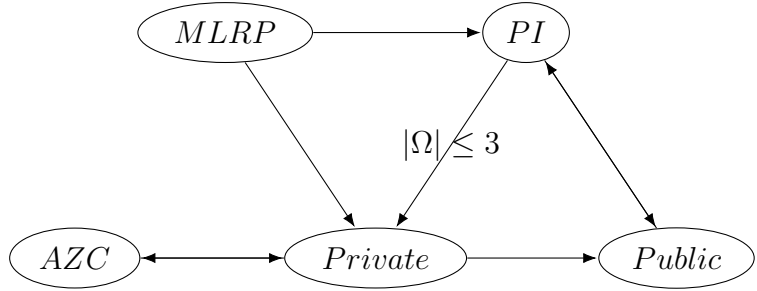
\begin{center}

	\begin{tikzpicture}
	\usetikzlibrary{shapes,decorations,arrows,calc,arrows.meta,fit,positioning}
\tikzset{
	-Latex,auto,node distance =1 cm and 1 cm,semithick,
	state/.style ={ellipse, draw, minimum width = 0.7 cm},
	point/.style = {circle, draw, inner sep=0.04cm,fill,node contents={}},
	bidirected/.style={Latex-Latex,dashed},
	el/.style = {inner sep=2pt, align=left, sloped}
}
	\node[state] (x) at (2,3) {$PI$};
	
	\node[state] (w) at (-4,0) {$AZC$};
	\node[state] (y) at (0,0) {$Private$};
	\node[state] (z) at (4,0) {$Public$};
	\node[state] (v) at (-2,3) {$MLRP$};
	
	\path (x) edge (y) node at (1,1.5) {$|\Omega|\leq 3$};
	\path (x) edge (z);
	\path (v) edge (x);
	\path (v) edge (y);
	\path (z) edge (x);
	\path (y) edge (z);
	\path (w) edge (y);
	\path (y) edge (w);
\end{tikzpicture}

\end{center}

Here $PI$ denotes that the information structure is pairwise informative; $Public$ denotes that the market gets learning when the signal is public; and $Private$ denotes that the market gets learning when the signal is private; $MLRP$ denotes that the information structure satisfies the monotonic likelihood ratio property; and $AZC$ denotes that the information structure satisfies Avery and Zemsky Condition. The arrow with $|\Omega|\leq 3$ means when the number of states is no greater than 3, pairwise informativeness implies market learning in the private signal case.

We present related literature and their connection with this paper in Section 2. The 3rd section presents the basic setup of model and information inference process. Section 4 characterizes the asymptotic learning conditions for a market with public signals and a market with private signals respectively. Also these learning conditions will be compared under different settings. The 5th section is Main Results where the sufficiency and necessity of the learning conditions will be proved. The 6th section is Discussion and Extensions where the relationship of AZC and PI under some prevalent conditions such as Monotonic Likelihood Ratio Property are discussed. Most proofs are in Appendix Omitted Proofs.

\section{Literature Review}
Extensive literature surveys on social learning are in \cite{bikhchandani2021information}. Early researchers have noticed that markets do not operate as desired because agents do not take actions concurrently once a market opens. \cite{banerjee1992simple}, \cite{bikhchandani1992theory} proposed the concept of “information cascades” as a learning outcome. They argue that in a sequential decision model, even rational agents are influenced by previous actions, and possibly ignoring the private information under some conditions. In their definition, once the agents ignore their own private information and make decisions solely based on history information, an informational cascade occurs. Then the market’s information aggregation is blocked and they enter a Bayesian equilibrium.

It follows by a classic question: when can people learn the truth and take optimal actions in a learning process? \cite{smith2000pathological} shows that unbounded belief is both sufficient and necessary to ensure learning when the number of states are finite. And for single crossing differences preferences, \cite{kartik2021observational} obtain directionally unbounded belief as a weaker learning condition than unbounded belief. When the decision makers are none-Bayesian due to computational cost, \cite{arieli2021sequential} shows that naive learning only requires the signal structure to be informative, while quasi-Bayesian learning is not assured correct learning.

There are also rich literature surveys in which learning outcomes arise under misspecification. \cite{frick2020misinterpreting} discuss the learning process when agents misinterpret others' preferences. \cite{bohren2021learning} allow both misspecification on distribution of other agents' type and signal structure, but their preferences are aligned to some extent such that confounded learning is avoided. These two paper still assume for common prior, while \cite{nyarko1991learning} and \cite{fudenberg2017active} consider misspecified prior beliefs.

The learning conditions discussed in the literature above are in a market without price, or the payoff only depends on the state and action. But the learning condition could be even weaker when the price is determined dynamically. \cite{avery1998multidimensional} simplify the trading model developed by \cite{glosten1985bid} where the price of the market is endogenous. They show that informational cascades are impossible when the prices incorporate all history information. The no informational cascade condition is necessary for asymptotic learning because if an informational cascade occurs before the belief reaches the truth, no new information can flow into the market which prevents learning. A key assumption over the information structure in \cite{avery1998multidimensional} ensures asymptotic learning, and the necessity of which will be supplemented by this paper. However, if transaction cost is taken into account, an informational cascade can occur before market learning. \cite{romano2007learning} shows that information asymmetry decreases as informative traders buy or sell, thus enter an information cascades set and stop learning. When it comes to a two-firms market where consumers do not know which product is better, \cite{arieli2019implications} shows that the better product will asymptotically prevail under observational learning as long as the signal structure satisfies vanishing margins condition. Without vanishing margins, the market may enter an wrong informational cascade with a positive probability, i.e., everyone buys the inferior product even when the better one is free.

\section{Model}
\subsection{Basics}
\textbf{Prior Belief:}\\
Consider a security market with infinite agents and market makers. The value of holding a share of the security, $\omega$, is drawn from a bounded\footnote{The boundedness of state space may not be necessary for the main results of this paper, but can greatly simplify the proofs.} set $\Omega$. For example let $\Omega=\{1,2,3\}$, then the value of the security could only be either 1, 2 or 3. The agents and market makers do not observe $\omega$ directly, but they share a prior belief over $\omega$, $\mu_0\in\Delta\Omega$. Formally speaking, at the beginning, state $\omega$ is a random variable defined on probability space $(\Omega,\mathcal{F},\mu_0)$ where $\mu_0$ is the common prior. $\mu_0(\omega)$ represents the probability of state $\omega$ in the prior belief. Sometimes when states are finite, we use a vector to denote the distribution over each state in agents' belief. For example, suppose $\mu_0=(1/3,1/3,1/3)$, then people initially believe that the value of the security is randomly sampled from $\{1,2,3\}$ with equal possibility. For convenience, the prior belief $\mu_0$ is supposed to has full support, i.e. $\mu_0(\omega)>0$ for all $\omega\in\Omega$. So that no state can be excluded at the beginning. By Bayes Rule, if $\mu_0(\omega^*)=0$ for some $\omega^*\in\Omega$, then $\mu_0(\omega^*|I)=0$ for any information $I\in\mathcal{F}$, hence introducing $\omega^*$ makes no sense.\\

Why can we assume common prior? If their priors are heterogeneous, they will soon trade until reach an agreement. Consider any pair of rational agents who value a stock at difference values, agent 1 values the stock at 1 dollar and agent 2 values the stock at 2 dollars. Agent 1 shall be happy to sell stocks to agent 2 at some price between 1 and 2 dollars. As they trade, one possibility is that either agent 1 exhausts her stocks or agent 2 exhausts her cash/credit. Another possibility is that during the trade process, both agent 1 increases her valuation because she thinks agent 2 may own some good news about the stock, also agent 2 decreases her valuation for symmetric reason. In the equilibrium, the remaining agents have homogeneous beliefs.\\

\noindent \textbf{Signal Structure:}\\
There are two types of agents: informative agents and noise agents. The probability of being an informative agent is $\eta$ and probability of being a noise agent is $1-\eta$, which is a common information for every participants. The type of each agent is unobservable to other agents. Each informative agent would receive an i.i.d. signal $s$ from the universe signal set $\mathbb{S}$, conditional on the state $\omega$. For example, in a normal information structure, typically assume $s|\omega\sim \mathcal{N}(\omega,\sigma^2)$, the signals are normally distributed with mean $\omega$ and a known variance $\sigma^2$. The signal space $\mathbb{S}$ is assumed to be a subset of real number, i.e. $\mathbb{S}\subseteq\mathbb{R}$. Albeit in a financial market the signals could be textual, e.g. the financial statement and meeting records, there always exists a real number that corresponds to textual information. The information structure is defined in probability space $(\Omega\times\mathbb{S},\mathcal{F}_{\Omega\times\mathbb{S}},f)$ where $f(s|\omega)$ denotes the probability density of signal $s$ given state $\omega$. Following \cite{kartik2021observational}, we assume that no signal can exclude any state, i.e. $f(s|\omega)>0$ for all $s$ and $\omega$. Also we assume $f(s|\omega)<\infty$ so that we do not need to deal with the problem involving $\infty/\infty$.\\

\noindent \textbf{Decision Making:}\\
At the beginning of date $t$, an agent will take the action $a_t\in A$. The action space $A=\{B,S,NT\}$ has 3 possible actions where $B$ represents buying a share, $S$ represents selling a share and $NT$ represents that no trade happens.  How do they decide their action choices? The informative agents are assumed to be rational and risk-neutral so that they choose their actions by maximizing their expected utility conditional on all observable information. We can derive their choice function. That is, they will choose to buy if their expected value is higher than the ask price, sell if lower than the bid price and otherwise no trade happens. \footnote{ In reality, even if a signals is not strong enough for a informative trader to trade, she still has incentive to post a new pair of bid ask prices that reveals some information about the signal. This paper leaves the discussion over ``informative market maker" for future research.} As for noise agents, they will buy, sell or do not trade with equal probability.\\

\noindent \textbf{Price setting:}\\
The market makers set a ask price $ask_t$ and a bid price $bid_t$ at the beginning of each date such that all market makers earn zero expected profit. All these prices are selected from a convex hull of $\Omega$ by market makers. Intuitively, market makers lose money to informative agents and earn money from noise agents on expectation because of inverse selection. The price $p_t$ is referred as the realized transaction price by date $t$, i.e., $p_{t+1}=ask_{t}$ if $a_t=B$ and $p_{t+1}=bid_{t}$ if $a_t=S$ for $t=0,1,2,\cdots$\footnote{$p_0$ does not exists in the market, however, it can be regarded as the prior expectation over the value for convenience.}.\\

\noindent \textbf{Payoffs:}\\
The payoff function for informative agents is as follows:
$$u(a, \omega, p)=(\omega-p){\mathbf{1}\{a=B\}}+(p-\omega){\mathbf{1}\{a=S\}}$$
As counterparties, the payoff function of market makers is just the opposite number of informative agents' payoff. The noise agents' strategy is independent to price $p_t$ and value $\omega$, so their payoff function does not matter. This model is a simplified version of Avery and Zemsky's model, because it implicitly assumed that everyone knows their own type (informative agent or noise agent). And it is more relax than the model in \cite{park2011herding} because we allow the state space $\Omega$ to be bounded instead of finite.

\subsection{Model Inference}
\hypertarget{modelinf}{}
\textbf{Public Signal Case:}\\
	At each date $t\geq 1$, an informative agent comes to the market, there is a signal $s_t$ open to the public in the public signal case. Now since information is perfectly revealed, inference of true state is simple. Recall that the existence of bid-ask spread is due to the information asymmetry\citep{glosten1985bid}. The price set by market makers is: $p_t=E[\omega|s_1,s_2,\cdots,s_t]$ and $p_0=E[\omega]$ where $s_t$ is i.i.d. given $\omega$. The informative agents will not trade because their expected values are the same as the market price, $p_t$. Now each market maker receives a zero profit in expectation, new competitors have no incentive to join the market. The market is then in a weakly perfect bayesion equilibrium (WPBE). Define the public belief $\mu_t\in\Delta\Omega$ to be a natural extension of prior $\mu_0$ such that it is the belief of market makers at date $t$ or the belief of an informative agent at date $t$ but before she observes any signal. 
Note that by observing signals, the market participants are able to distinguish whether an agent is informative or noise. Once the previous agent is noise, they simply inherit the public belief of previous period, i.e. $$\mu_t(\omega)=\mu_{t-1}(\omega)$$ 
Upon receive a signal, public belief updates by Bayes rule as follows:
$$\mu_t(\omega)=\mu_{t-1}(\omega)\cdot \frac{f(s_t|\omega)}{\sum_{w\in\Omega}\mu_{t-1}(w)f(s_t|w)}$$
In the public signal case, public beliefs follow the above law of motion.\\

\noindent \textbf{Private Signal Case:}\\
	Now assume signals are private to each trader and the market maker only observe the action history. Let $H_t=\{a_1,\cdots,a_{t-1}\}$ denote the history of actions by date $t$.\footnote[1]{Let $H_1=\phi$ such that $E[\omega|H_1]=E[\omega]$}
	We do not include the history prices in $H_t$ as \cite{park2011herding} did because prices are determined by prior belief and history actions. There is no difference whether we include the price or not in terms of information. 
	
	If a trader is informative at date $t$, her strategy is as follows:
	\begin{equation}
		a_t=
		\begin{cases}
			B & \text{if } E[\omega|s_t,H_t]>ask_t\\
			S & \text{if } E[\omega|s_t,H_t]<bid_t\\
			NT & otherwise
		\end{cases}	
	\end{equation}
She buys when her subjective value is higher than ask price and sells when her subjective value is lower than bid price. Otherwise she would rather stay still.
At each date $t$, given the agent is informative there exists a signal partition $\{S^B_t,S^S_t,S^{NT}_t\}$ given history $H_t$ and the prior $\mu_{0}$, such that the following conditions are satisfied:\\
\begin{itemize}
	\item $S^B_t=\{s_t:E[\omega|s_t,H_t]>ask_t\}$
	\item $S^S_t=\{s_t:E[\omega|s_t,H_t]<bid_t\}$
	\item $S^{NT}_t=\{s_t:bid_t\leq E[\omega|s_t,H_t]\leq ask_t]\}$
\end{itemize}
Recall that the agent could also be noise agent with probability $\eta$.
	Using the Bayes Rule and the signal partition above, the public belief updates as follows:
	\begin{align*}
	\mu_{t}(\omega)&=\mu_{t-1}(\omega)\cdot \frac{f(a_{t}|\omega,H_{t})}{\sum_{w\in\Omega}\mu_{t-1}(w)f(a_{t}|w,H_{t})}
	\end{align*}
where $f(a_{t}|\omega,H_{t})=\eta/3+(1-\eta)f(S^{a_{t}}_{t}|\omega)$ and $t\geq 1$.
The informative trader's private belief updates as follows:
$$\mu_t(\omega|s_t)=\mu_t(\omega)\cdot\frac{f(s_t|\omega)}{\sum_{w\in\Omega}\mu_t(\omega)f(s_t|\omega)}$$

Although the private signal is unobservable to other market participants, they know that the signal belongs to which part of the signal space.

Under the equilibrium, market makers earn zero expected profit whether an agent comes to buy or sell. The ask price posted at date $t$ satisfies the following equation:
	$$\frac{\eta}{3}\big(ask_t-E[\omega|H_t]\big)=(1-\eta)\cdot \int_{s\in S_t^B}\big(E[\omega|s,H_t]-ask_t\big)f(s|H_t)ds$$
	where $f(s|H_t)=\int_{\omega\in\Omega}f(s|\omega)\mu_t(\omega)d\omega$. That is, $f(s|H_t)$ is a subjective signal distribution conditional on what agents observed. The left hand side is the product of probability of a noise agent comes to buy and the expected profit market makers earn from the noise agent; the right hand side is the product of probability of an informative agent comes and the expected loss market makers suffer from this informative agent's buy action. Similarly we are able to construct the equation for bid price at each date.
	As for the existence uniqueness of their solutions, \cite{avery1998multidimensional} has shown in their proposition 1 by continuity and monotonicity.

	\begin{lemma} 
		\hypertarget{lem1}{}
		At date $t$, if there is no trade at date $t$, the price will not change; if the trader buys, the price will rise; if the trader sells, the price will fall.
	\end{lemma}

	This lemma is quite intuitive because in our model, an informative agent only buys (sells) when her posterior expectation after observing the signal is strictly higher (lower) than the current ask (bid) price. This result simplifies future discussions, because we can ignore the period of no trade when discussing the properties of price sequence.

To summarize the model, we draw the following timeline,

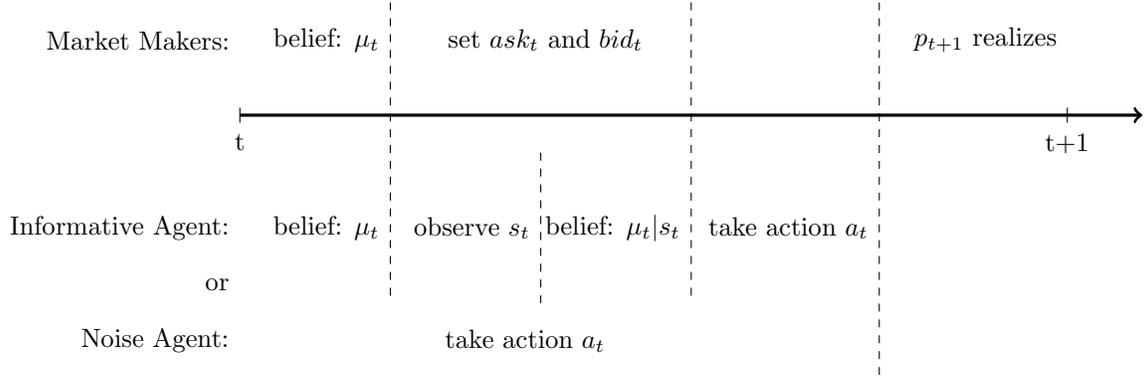
\begin{figure}
    \centering

\begin{tikzpicture}[y=1cm, x=1cm,  font=\fontsize{10}{10}\selectfont]    

\draw[line width=1.2pt, ->](0,0) -- coordinate (x axis) (12,0) ; 
\draw (0,0.1) -- (0,-0.1) node[below] {t};

\draw (11,0.1) -- (11,-0.1) node[below] {t+1};
\draw[dashed] (2,1.5) -- (2,-2.5);
\draw[dashed] (6,1.5) -- (6,-2.5);
\draw[dashed] (8.5,1.5) -- (8.5,-3.5);
\draw[dashed] (4,-0.5) -- (4,-2.5);

\draw (0,1) node[left] {Market Makers:};
\draw (2,1) node[left] {belief: $\mu_t$};
\draw (5.5,1) node[left] {set $ask_t$ and $bid_t$};
\draw (11,1) node[left] {$p_{t+1}$ realizes};

\draw (0,-1.5) node[left] {Informative Agent:};
\draw (2,-1.5) node[left] {belief: $\mu_t$};
\draw (4,-1.5) node[left] {observe $s_t$};
\draw (6,-1.5) node[left] {belief: $\mu_t|s_t$};
\draw (8.5,-1.5) node[left] {take action $a_t$};

\draw (0,-2.25) node[left] {or};

\draw (0,-3) node[left] {Noise Agent:};
\draw (5,-3) node[left] {take action $a_t$};

\end{tikzpicture}
\caption{The timeline of market participants' believes, strategies and price during period $t$.}
\end{figure}

\newpage
	\section{Conditions for Market Learning}
	This section discusses further details in the markets with public signals and private signals as well as their learning conditions PI and AZC. In addition, the proof for the sufficiency and necessity of these conditions will be presented. We first give the formal definition for market learning, pairwise informativeness and Avery and Zemsky condition. Then show their relationship under private/public signal cases.
	
		\begin{definition}
		A market gets learning if starting from any full support prior $\mu_0$, the public belief over the true value $\mu_t(\omega)$ converges to 1 in probability, i.e. $\lim\limits_{t\rightarrow\infty}\mu_t(\omega)=1$, where $\omega$ is the true value of the asset.
	\end{definition}
	In fact, the definition of market learning can also express in terms of price convergence, i.e. for all full support prior $\mu_0$, $$\lim\limits_{t\rightarrow\infty}p_t=\omega$$
	The two definitions are equivalent, because the convergence holds for all full support priors. It is possible that for some prior, the price converges to the true value while the public belief does not converge to 1 on the true value. But if we disturb the prior a little bit, the price will then converge to somewhere else.

	\subsection{Pairwise Informativeness}
    \hypertarget{pi}{}
	\begin{definition}
	A signal structure is 
	{pairwise informative} if $\:\forall w_1\neq w_2\in\Omega$, there exists a positive measure of signal $S_{\omega_1,\omega_2}\subseteq \mathbb{S}$ such that for all $s_{\omega_1,\omega_2}\in S_{\omega_1,\omega_2}$: $$f(s_{\omega_1,\omega_2}|\omega_1)\neq f(s_{\omega_1,\omega_2}|\omega_2)$$
	where $f(s|\omega)$ represents the probability density of signal $s$ appears given the state is $\omega$.
    \end{definition}
When a signal structure is PI (pairwise informative), then each pair of states are identifiable, i.e., one can always identify a state given its signal distribution. In other words, each state generates signals with unique distribution. This condition reduces to so-called ``informativeness" in the tradition of observational learning model under $|\Omega|=2$ which is the weakest condition in all learning models.

\begin{lemma} 
	\hypertarget{lem2}{}
	If a signal structure is pairwise informative, then for all $\omega_1\neq\omega_2\in\Omega$, there exists $s_1, s_2\in\mathbb{S}$ such that:
	$$f(s_1|\omega_1)>f(s_1|\omega_2)$$
	and
	$$f(s_2|\omega_1)<f(s_2|\omega_2)$$

\end{lemma}
If a state has higher probability to generate one signal than another state, it must be dominated at another signal. Because whatever the state is, the sum of probability of signal is always 1.\\
Suppose not, i.e. there exists $\omega_1\neq\omega_2\in\Omega$ such that $f(s|\omega_1)\geq f(s|\omega_2)$  for all $s\in\mathbb{S}$ and $f(s|\omega_1)> f(s|\omega_2)$ for a positive measure of $s\in\mathbb{S}$. Then take the integration on both sides,
$$1=\int_{s\in\mathbb{S}}f(s|\omega_1)>\int_{s\in\mathbb{S}}f(s|\omega_2)=1$$
which is a contradiction.

	\begin{prop}
	\hypertarget{p1}{} 
		  The market with public signal gets learning if and only if the signal structure is pairwise informative .\\
	\end{prop}
Detailed proof for the sufficiency and necessity of proposition is in appendix. As long as there are enough i.i.d signals, we can estimate the empirical distribution of signal and compare with the signal structure. By the law of large number, the empirical distribution function will converge to the true distribution function as time goes to infinity.

\subsection{Avery and Zemsky's Condition}
A typical learning condition is first proposed by \cite{avery1998multidimensional}, hereafter we call it \hypertarget{az}{AZC}. It only requires boundedness on state space while no restrictions on the size of it. However, Avery and Zemsky only provided the intuition that AZC is sufficient for long run learning. This paper will complete the formal proof for AZC's sufficiency and give the proof for AZC's necessity.\\
\begin{definition}
	A signal structure satisfies AZC if for all full support belief $\mu\in\Delta\omega$ and $\delta>0$ such that $|E_\mu[\omega]-\omega|> \delta$, there exists $\epsilon>0$ and $s\in\mathbb{S}$ such that $|E_\mu[\omega|s]-E_\mu[\omega]|>\epsilon$.
\end{definition}
This definition has some differences comparing to the definition in \cite{avery1998multidimensional}. Here we remove the history information in the expressions and add the quantifier ``for all full support belief". This modification makes the definition briefer and understandable. Basically, AZC says if there is a difference between the price and the true state, then there exists a signal (with positive probability) that moves the price by an amount uniformly bounded from 0. It is not equivalent to say for all full support $\mu$, if $E_\mu [\omega]\neq \omega$, then there exists a signal $s\in\mathbb{S}$ such that $E_\mu[\omega|s]\neq E_\mu[\omega]$. Because then we can construct a sequence such that for some prior belief $\mu_1$ and signal structure $f$, even if ``best"\footnote{ ``Best" signal means the signal that would help the private posterior move towards the true state most.} signal is observed in each period, $E_{\mu_t}[\omega]=0.8\omega^*-0.5^t$ where $\omega^*$ is the true state. The movement of public belief is always positive, but the public belief never converges to the truth! 

To fixed the problem above, one solution is to substitute ``for all full support beliefs" with ``for all stochastic beliefs", i.e., for all $\mu\in\Delta\Omega$ and $\omega\in\Omega$ such that $\mu(\omega)<1$, there exists a signal $s\in\mathbb{S}$ such that $E_\mu[\omega|s]\neq E_\mu[\omega]$. Without the full support requirement, the public belief can be placed on any pair of states (and 0 on other states in the state space). This protects the correct state from being confounded with any other state.

\section{Main Results}
On the one hand, in a market with public signals, \hyperlink{pi}{pairwise informativeness} is both sufficient and necessary for market learning. On the other hand, in a market with private signals, pairwise informativeness is necessary for market learning; and when the number of states, $|\Omega|$, is no greater than 3, then pairwise informativeness is sufficient. Moreover, \hyperlink{az}{AZC} is both sufficient and necessary in a market with private signals when state space $\Omega$ is bounded.

\subsection{Sufficiency of AZC}
In the Proposition 4 of \cite{avery1998multidimensional}, they provided an intuition of why market learning is assured under AZC in a private signal market. In this section, a more detailed reasoning of the sufficiency of AZC will be presented.

First prove the existence of the price limit. The history $H_t$ is a filtration, i.e. $H_s\subset H_t$ for $1<s<t<\infty$, then $E[p_{t+1}|H_t]=E[E[\omega|H_{t+1}]|H_t]=E[\omega|H_t]=p_t$. So the price is a martingale with respect to $H_t$, and has bounded expectation $E[|p_t|]<\infty$. By the Martingale Convergence Theorem, $lim_{t\rightarrow\infty}p_t$ exists almost surely.

Then show the price limit is the true value by contradiction. Suppose the true value of the asset is $\omega^*$, instead of $\lim\limits_{t\rightarrow\infty}p_t=\omega'$. Take any $\delta\in(0,|\omega^*-\omega'|)$, by the definition of convergence in probability, for all $\xi>0$, there exists $T\in\mathbb{N}$ such that 
$$\forall t>T, \: Pr(|p_t-\omega'|<|\omega^*-\omega'|-\delta)>1-\xi$$
By the absolute value inequality, $|p_t-\omega^*|\geq|\omega'-\omega^*|-|p_t-\omega'|$. Then
\begin{align*}
	Pr(|p_t-\omega^*|>\delta)&\geq Pr(|\omega'-\omega^*|-|p_t-\omega'|>\delta)\\
	&=Pr(|p_t-\omega'|<|\omega'-\omega^*|-\delta)\\
	&>1-\xi
\end{align*}
By choosing an arbitrarily small $\xi$, we are arbitrarily certain about $|E[\omega|H_t]-\omega^*|>\delta$ for $t>T$.

It implies that there exists a signal $s$ and a constant $\epsilon$ such that $|E[\omega|s,H_t]-E[\omega|H_t]|>\epsilon$ for almost all $t>T$ by \hyperlink{az}{AZC}. The trader who observes $s$ will either buy or sell, $s\in S^B_t\cup S^S_t$, for almost all $t>T$.

Then by Markov inequality, when $s\in S^B_t$ and s is observed at date $t$,
\begin{align*}
	E\big[E[\omega|s_t,H_t]-p_t\big|a_t=B,H_t\big] &\geq \epsilon Pr(E[\omega|s_t,H_t]-p_t\geq\epsilon \big|a_t=B,H_t)\\
	p_{t+1}-p_{t}&\geq \epsilon Pr(s_t=s|\omega=\omega^*)
\end{align*}

Similarly, we can show that when $s\in S^S_t$, $p_t-p_{t+1}\geq \epsilon P(s_t=s|\omega=\omega^*)$.
Hence $|p_{t+1}-p_t|\geq \epsilon Pr(s_t=s|\omega=\omega^*)$ for some $t>T$, where the right hand side is a constant. In other words, the price will move by at least $\epsilon Pr(s_t=s|\omega=\omega^*)$, with a non-trivial probability which is at least $(1-\xi)Pr(s_t=s|\omega=\omega^*)$ for any date $t>T$. This contradicts the definition of convergence in probability.

AZC is inspired from the minimal ``useful" information assumption from the setup of Avery and Zemsky (1998). It is a general condition regardless of the number of states. It only requires a bounded state space to ensure the application of Martingale Convergence Theorem. Avery and Zemsky (1998) gave an illustrative proof in their Proposition 4, which provides the main idea of this proof. Additionally, we modify the quantifiers of AZC and finally derive the formal math proof. In the model of Park and Sabourian (2011), there is a similar assumption that in any finite history $H_t$, there is always a signal $s\in\mathbb{S}$ such that $E[\omega|s,H_t]\neq E[\omega|H_t]$. It is a weaker assumption than AZC, but no one has given the sufficiency proof.

\subsection{Necessity of AZC}
The necessity of \hyperlink{az}{AZC} for market learning has not been discussed in the previous research. What if AZC does not hold? This section shows that AZC is also necessary for market learning in private signal case.
\begin{definition}
	An informational cascade occurs at date $t$ if for all $s\in\mathbb{S}$, $E[\omega|s,H_t]=E[\omega|H_t]$.
\end{definition}
The negation of AZC should be:\\
There exists a full support belief $\mu\in\Delta\omega$ and $\delta>0$ such that $|E_\mu[\omega]-\omega|>\delta$, for all $\epsilon>0$ and $s\in\mathbb{S}$, $|E_\mu[\omega|s]-E_\mu[\omega]|\leq\epsilon$. That is, for all $s\in \mathbb{S}$, $E_\mu[\omega|s]=E_\mu[\omega]$.\\
By the definition of informational cascade, the negation of AZC can be stated as:\\
There exists a full support prior $\mu_0\in\Delta\omega$ and a finite history $H_t^*\subset A^{t-1}$ and $\delta>0$ such that $|E_\mu[\omega]-\omega|>\delta$, an informational cascade occurs at date $t$. \\
Once an informational cascade occurs, there is no possibility for price to move. That is, $E[\omega|H_s]=E[\omega|H_t^*]$ for all $H_t^*\subseteq H_s$. Suppose $\xi=Pr(H_t=H_t^*)>0$, then $$Pr(|E[\omega|H_t]-\omega|>\delta)\geq\xi$$
And for all $s>t$, $$Pr(|E[\omega|H_s]-\omega|>\delta)\geq Pr(|E[\omega|H_s]-\omega|>\delta,H_t=H_t^*)\geq \xi$$
It means that starting with some prior $\mu_0$, $p_t\nrightarrow \omega$ with some non-trivial probability $\xi$, which contradicts to the definition of market learning. Hence the market cannot get learning without AZC.

\subsection{Sufficiency of PI under 3 States}
In this section, we will show that if $|\Omega|\leq 3$ and PI holds, then the market in the private signal case also gets learning. The proof will focus on the properties of limit belief. When the number of states exceeds 3, we will provide a counter-example where PI is no longer sufficient for learning in the private signal case.

\begin{lemma}
\hypertarget{lem3}{}
The limit belief $\tilde{\mu}$ exists, i.e., 
$$\mu_t(\omega)\xrightarrow{a.s.}\tilde{\mu}(\omega)$$
\end{lemma}

\hypertarget{ntc}{}
\begin{definition}
	No Trade Condition:
	under the limit belief $\tilde{\mu}$, there is an informational cascade, i.e., $E_{\tilde{\mu}}[\omega]=E_{\tilde{\mu}}[\omega|s]\: \forall s\in\mathbb{S}$. Hence informative agents will stop trade.
\end{definition}
The No Trade Condition is a natural implication from \hyperlink{lem1}{Lemma 1}. Because if the limit belief does not lie in the region of informational cascade, there is a positive probability that the belief and the price will move by some extent bounded away from 0.

\begin{lemma}
\hypertarget{lem4}
Under pairwise informativeness, when $|\Omega|\leq 3$, for all $\omega\in\Omega$, $\tilde{\mu}(\omega)\in\{0,1\}$.
\end{lemma}

\begin{lemma}
\hypertarget{lem5}
If for all full support prior $\mu_0$, $\tilde{\mu}(\omega)\in\{0,1\}$, then $\omega^*$ is the true value of the asset if $\tilde{\mu}(\omega^*)=1$.
\end{lemma}

The intuition for Lemma 5 is that standard Bayesian learner's belief always has support on correct state. Assume $\omega^*$ is the true state or the true value of the trading asset and $\neg \omega^*$ be the set incorrect states. Let $\lambda_t$ be the likelihood ratio of public belief over the incorrect states over the public belief over the true state at date $t$, i.e.,
$$\lambda_t = \frac{\mu_t(\neg\omega^*)}{\mu_t(\omega^*)}$$
Note that $\mu_t(\neg\omega^*)$ is a abbreviation of $\mu_t(\omega\in\neg\omega^*)$, the probability of incorrect state measured by public belief at date $t$. If we take the conditional expectation over $\lambda_{t+1}$ conditional on $\lambda_t$
Then $\lambda_t$ is a martingale, because
\begin{align*}
    E[\lambda_{t+1}|\lambda_t]&=\sum\limits_{a_{t}\in A}f(a_t|\omega^*,H_t)\cdot \lambda_{t+1}\\
    &=\sum\limits_{a_{t}\in A}f(a_t|\omega^*,H_t)\cdot \lambda_t\cdot\frac{f(a_t|\neg\omega^*,H_t)}{f(a_t|\omega^*,H_t)}\\
    &=\sum\limits_{a_{t}\in A}f(a_t|\neg\omega^*,H_t)\cdot \lambda_t\\
    &=\lambda_t
\end{align*}
And $\lambda_t$ is $L^1$ bounded, because $E|\lambda_t|=E[\lambda_t]=E[\lambda_1]=\frac{1-\mu_1(\omega^*)}{\mu_1(\omega^*)}<\infty$.
By Martingale Convergence Theorem, $\lambda_\infty<\infty$ exists, that is, $\tilde{\mu}(\omega^*)\neq0$. Then the limit belief over the true state can only be 1.

Obviously, if we combine the Lemma 3, 4 and 5, we get the following theorem:\\
\begin{thm}
When the number of states is no greater than 3 and the information structure is pairwise informative, the market gets learning in the private signal case.
\end{thm}
Although AZC has been shown to be both sufficient and necessary for market learning in the private signal case, it is difficult to verify even when there are only a few states. While PI is a simple condition and can be verified quickly (a few seconds with human brain) when there are only 2 or 3 states.

When there are at least 4 states, PI is no longer a sufficient condition for market learning in the private signal case. Here is a counter-example:

\begin{example}
There are four states with uniform prior, i.e., $\Omega=\{0,1,2,3\}$ and $\mu(\omega)=0.25$ for all $\omega\in\Omega$. And there are four signals in signal space, $\mathbb{S}=\{s_1,s_2,s_3,s_4\}$. The signal structure is characterized in \hyperlink{table1}{Table 1}.
\end{example}
\begin{table}
    \centering
    \begin{tabular}{c|cccc}
    $f(s|\omega)$  & $s_1$ & $s_2$ & $s_3$ & $s_4$ \\
    \hline
      0  & 0.3 & 0.2 & 0.2 & 0.3 \\
      1  & 0.1 & 0.4 & 0.3 & 0.2 \\
      2  & 0.4 & 0.1 & 0.3 & 0.2 \\
      3  & 0.2 & 0.3 & 0.2 & 0.3
    \end{tabular}
    \caption{Signal Structure of Example 5.1\hypertarget{table1}{}. The $(i,j)$ element represents the probability of signal $s_j$ given state $i-1$.}
\end{table}
It is convenient to verify that the signal structure above is pairwise informative, i.e., no two states generate the same signal distribution.  In this 4-state example, matrix operation simplifies our notations, so we will use $\mu$ and $\mu_s$ to represent vectors of $\mu(\omega)$ and $\mu(\omega|s)$. Here
$$\mu=(0.25, 0.25, 0.25, 0.25)$$
represents the uniform prior vector.
Recall how we calculate posterior beliefs after observing each signal by Bayes rule.
$$\mu(\omega|s)=\frac{\mu(\omega)f(s|\omega)}{\sum_{w\in\Omega}\mu(w)f(s|w)}$$
Then posterior vectors under each signal are as follows:
$$\mu_{s_1}=(0.3,0.1,0.4,0.2)$$
$$\mu_{s_2}=(0.2,0.4,0.1,0.3)$$
$$\mu_{s_3}=(0.2,0.3,0.3,0.2)$$
$$\mu_{s_4}=(0.3,0.2,0.2,0.3)$$
Based on the prior belief and the posteriors, we then calculate the prior expectation and posterior expectations:
$$\mathbb{E}[\omega]=\mathbb{E}[\omega|s_1]=\mathbb{E}[\omega|s_2]=\mathbb{E}[\omega|s_3]=\mathbb{E}[\omega|s_4]=1.5$$
which means that informative agents will not trade regardless of signals they receive. And the subsequent agents will not benefit from observing informative agents' behaviors. The learning stops and an informational cascade occurs immediately. Similar examples can be constructed when there are more than 4 states. This example tells us that PI is no longer sufficient for market learning when there are at least 4 states.
\section{Discussion and Extension}\hypertarget{s6}{}
\subsection{Relationship between public and private signal case}
What is the relationship between the public signal case and the private signal case? It is intuitive that if a market gets learning with private signal, then it should also get learning with public signal. Because the information from the private signal case is actually a subset of the public signal case given the prior belief, i.e.
$$H_t\subseteq\{s_1,s_2,\cdots,s_{t-1}\} |\mu_0 \: \forall t>0$$
An agent with prior $\mu_0$ can derive history information $H_t$ in private signal case based on $\{s_1, s_2,\cdots,s_{t-1}\}$.
Hence when $\omega^*$ is the true value of the asset, and if $E[\omega|H_t]\xrightarrow{p}\omega^*$,
then
$E[\omega|s_1,s_2,\cdots,s_{t-1}]\xrightarrow{p}\omega^*$.\\
Suppose not, e.g. $E[\omega|s_1,s_2,\cdots,s_{t-1}]\xrightarrow{p}\omega'\neq\omega^*$, then
\begin{align*}
	\lim\limits_{t\rightarrow\infty}E[\omega|H_t]&=\lim\limits_{t\rightarrow\infty}E[E[\omega|s_1,s_2,\cdots,s_{t-1}]|H_t]\\
	&=\lim\limits_{t\rightarrow\infty}E[\omega'|H_t]\\
	&=\omega'\neq\omega^*
\end{align*}
which contradicts that $E[\omega|H_t]\xrightarrow{p}\omega^*$.

Then we can argue that when the number of states is no greater than 3, PI is also necessary for market learning in the private signal case. Since PI is both sufficient and necessary for a market to get learning in the public signal case. And if a market gets learning with private signal, then it should also get learning with public signal.\\

\subsection{MLRP in Market Learning}
The monotonic likelihood ratio property (MLRP) established by \cite{brown1976complete} is a strong but widely used condition on the type distribution of agents in mechanism design. Most solutions to mechanism design models assume a type distribution to satisfy the MLRP. In this paper, we show how MLRP related to learning conditions like AZC and PI.
\begin{definition}
    An information structure has monotonic likelihood ratio property if for all $s_L<s_H$ and $\omega_L<\omega_H$
    $$f(s_L|\omega_L)f(s_H|\omega_H)\geq f(s_L|\omega_H)f(s_H|\omega_L)$$
    If the above inequality is strict, then strict MLRP holds.
\end{definition}

\begin{prop}
      Strict monotonic likelihood ratio property implies Pairwise Informativeness.
\end{prop}
\begin{proof}
The contrapositivity of this proposition is that PI does not hold implies that strict MLRP does not hold.\\
When PI does not hold, there exist $\omega_L<\omega_H$ such that for all $s\in S$,
$$f(s|\omega_L)=f(s|\omega_H)$$
Now take $s_L<s_H$, we have
$$\frac{f(s_L|\omega_L)}{f(s_L|\omega_H)}=\frac{f(s_H|\omega_L)}{f(s_H|\omega_H)}=1$$
which contradicts to the definition of strict MLRP. So strict MLRP does not hold and the contrapositivity of this proposition is true.
\end{proof}

So by proposition 2, strict MLRP is a stronger assumption than PI, it certainly implies learning in public signal case and when $|\Omega|\leq 3$, implies learning in private signal case as PI does. But what about the private signal case when $|\Omega|> 3$? This paper shows the answer is yes when $|\Omega|$ is bounded in the following.

\begin{lemma}
      Under monotonic likelihood ratio property and pairwise informativeness, $\tilde{\mu}(\omega)\in\{0,1\}$ for all $\omega\in\Omega$.
\end{lemma}
\begin{proof}
Suppose not, then the limit belief $\tilde{\mu}$ is supported on $\Omega_R\subseteq\Omega$ where $|\Omega_R|\geq 2$. That is,
$\tilde{\mu}(\omega)>0$
for all $\omega\in\Omega_R$.\\
Then by MLRP and PI, for all $\omega_L<\omega_H\in\Omega_R$ and $s_L<s_H\in S$,
$$\frac{\tilde{\mu}(\omega_H|s_H)}{\tilde{\mu}(\omega_H|s_L)}=\frac{f(s_H|\omega_H)}{f(s_L|\omega_H)}\geq\frac{f(s_H|\omega_L)}{f(s_L|\omega_L)}=\frac{\tilde{\mu}(\omega_L|s_H)}{\tilde{\mu}(\omega_L|s_L)}$$
The above inequality is strict for at least one group of $\omega_L<\omega_H$ and $s_L<s_H$.
By \cite{milgrom1981good}, it implies that the limit belief conditional on $s_H$ strictly First Order Stochastic Dominates (FOSD) that conditional on $s_L$. Then by the property of FOSD, $\omega$ is an increasing function of $\omega$, we have
$$\int\limits_{\omega\in\Omega_R}\omega\tilde{\mu}(\omega|s_H)d\omega>\int\limits_{\omega\in\Omega_R}\omega\tilde{\mu}(\omega|s_L)d\omega$$
$$E_{\tilde{\mu}}[\omega|s_H]>E_{\tilde{\mu}}[\omega|s_L]$$
Then either $E_{\tilde{\mu}}[\omega|s_H]\neq E_{\tilde{\mu}}[\omega]$ or $E_{\tilde{\mu}}[\omega|s_L]\neq E_{\tilde{\mu}}[\omega]$. So the \hyperlink{ntc}{No Trade Condition} is violated. Hence that for all $\omega\in\Omega$, $\tilde{\mu}(\omega)\in\{0,1\}$ is the only possibility.
\end{proof}

Now combine Lemma 3, 5 and 6, we find that under MLRP and PI, $\tilde{\mu}(\omega^*)=1$ where $\omega^*$ is the true value. That is, ``MLRP + PI" also implies market learning in the private signal case. And this sufficient learning condition does not restricted by the cardinality of state space. However, ``MLRP + PI" is not necessary for market learning.\\

\section{Conclusion}
We develop a general framework to study long run learning with competitive prices, which captures the impact of state space and whether signals are private or public on information-gathering process. A key contribution of our framework is the ability to connect the learning conditions of a market with private signals and a market with public signals. Our main result completes the sufficiency and necessity that asymptotic learning holds for private signal case if and only if Avery and Zemsky's Condition satisfies. This paper also shows the equivalence of AZC and the learning condition of public signal case, PI, when the number of state is no greater than 3. In addition, our result incorporates a prevalent information condition, MLRP, into the existing learning framework, and shows that ``MLRP $+$ PI" is a stronger condition than both AZC and PI for market learning with competitive prices.

\section{Acknowledgment}
The author does not have funding that relates to the research described in this paper. The author has no relevant or material financial interests that relate to the research described in this paper.

\newpage
\appendix
\section{Omitted Proofs}
\textbf{Proof for \hyperlink{lem1}{Lemma 1}:}
	\begin{proof}
		According to the signal partition, if there is no trade, the price at next date is $E[\omega|H_{t+1}]=E[\omega|a_t=NT,H_t]=E[\omega|s_t\in S^{NT}_t,H_t]=E[\omega|H_t]$;\\
		if the trader buys, the price at next date is $E[\omega|H_{t+1}]=E[\omega|a_t=B,H_t]=E[\omega|s_t\in S^{B}_t,H_t]>E[\omega|H_t]$;\\
		if the trader sells, the price at next date is $E[\omega|H_{t+1}]=E[\omega|a_t=S,H_t]=E[\omega|s_t\in S^{S}_t,H_t]<E[\omega|H_t]$.
	\end{proof}

\noindent \textbf{Proof for \hyperlink{p1}{Proposition 1}}
\begin{proof}
		\noindent\textbf{Sufficiency:}\\
		Suppose the true value of the asset is $\omega_0$. By the definition of pairwise informativeness,  for all $\omega_1\in\Omega\backslash\{\omega_0\}$, there exists a signal $s_{01}\in \mathbb{S}$ or a positive measure of signals $S_{01}\subset\mathbb{S}$ such that $f(S_{01}|\omega_0)\neq f(S_{01}|\omega_1)$. In the following discussion $S_{01}$ will be used to represent both cases and $S_{01}=\{s_{01}\}$ when signal space is discrete. By the Weak Law of Large Numbers, $$\frac{1}{T}\sum_{t=1}^{T}\mathbf{1}\{s_t\in S_{01}\}\xrightarrow{p}P(s_t\in S_{01})=f(S_{01}|\omega)$$
		 By Slutsky's theorem,
		$$\frac{|\frac{1}{T}\sum_{t=1}^{T}\mathbf{1}\{s_t\in S_{01}\}-f(S_{01}|\omega_0)|}{|\frac{1}{T}\sum_{t=1}^{T}\mathbf{1}\{s_t\in S_{01}\}-f(S_{01}|\omega_1)|}\xrightarrow{p}\frac{|f(S_{01}|\omega_0)-f(S_{01}|\omega_0)|}{|f(S_{01}|\omega_0)-f(S_{01}|\omega_1)|}=0$$
		for $T\rightarrow\infty$. The market maker is able to distinguish the true value $\omega_0$ from any other states in the limit. 
		
		\noindent\textbf{Necessity:}\\
		The negation of the condition becomes: $\exists \omega_1\neq\omega_2\in\Omega$ such that  for all signal subsets $S_{12}\subset\mathbb{S}$,
		$$f(S_{12}|\omega=w_1)= f(S_{12}|\omega=w_2).$$
		Then $\frac{\mu_t(\omega_1)}{\mu_t(\omega_2)}=\frac{\mu_{t-1}(\omega_1)}{\mu_{t-1}(\omega_2)}\frac{f(s_t|\omega_1)}{f(s_t|\omega_2)}=\frac{\mu_0(\omega_1)}{\mu_0(\omega_2)}$.
		As a result, the market will never distinguish $\omega_1$ and $\omega_2$ from the signal if one of them is the true state.
	\end{proof}

	\noindent\textbf{Proof for \hyperlink{lem3}{Lemma 3}}
	\begin{proof}

	Recall the dynamic process of the public belief:
\begin{align*}
	\mu_{t+1}(\omega)&=\mu_{t}(\omega)\cdot \frac{f(a_{t+1}|\omega,H_t)}{\sum_{w=0}^{2}\mu_{t}(w)f(a_{t+1}|w,H_t)}	\\
	& =\mu_{t}(\omega)\cdot \frac{f(S^{a_{t+1}}_{t+1}|\omega)}{\sum_{w=0}^{2}\mu_{t}(w)f(S^{a_{t+1}}_{t+1}|w)}	
\end{align*}
	Then take the conditional expectation with respect to $H_t$,
	\begin{align*}
		E[\mu_{t+1}(\omega)|H_t]&=\mu_{t}(\omega)\cdot E[\frac{f(S^{a_{t+1}}_{t+1}|\omega)}{\sum_{w=0}^{2}\mu_{t}(w)f(S^{a_{t+1}}_{t+1}|w)}|H_t]\\
		&=\mu_{t}(\omega)\cdot\sum_{a_{t+1}\in A}\Big\{\frac{f(S^{a_{t+1}}_{t+1}|\omega)}{\sum_{w=0}^{2}\mu_{t}(w)f(S^{a_{t+1}}_{t+1}|w)}\sum_{w=0}^{2}\mu_{t}(w)f(S^{a_{t+1}}_{t+1}|w)\Big\}\\
		&=\mu_{t}(\omega)
	\end{align*} 
Hence $\{\mu_{t}(\omega)\}_{t=0}^{\infty}$ is a martingale and is bounded in the sense that $$\sup\limits_{t\in\mathbb{N}, \omega\in\Omega}|\mu_{t}(\omega)|\leq1$$
By the Martingale Convergence Theorem, the limit of $\mu_{t}(\omega)$ must exist for all $\omega\in\Omega$. Let $\tilde{\mu}(\omega)$ denote the limit of $\mu_{t}(\omega)$, then
$\tilde{\mu}(\omega)=\tilde{\mu}(\omega|a)$ for all $a\in A$. However, under the limit belief $\tilde{\mu}$, the action could only be $NT$, otherwise by \hyperlink{lem1}{Lemma 1}, the price will move and hence the public belief $\tilde{\mu}(\omega)$ will change.
\end{proof}

\noindent\textbf{Proof for \hyperlink{lem4}{Lemma 4}}

\begin{proof}
Assume $\Omega = \{\omega_1,\omega_2,\omega_3\}$ be the state space. The case when $|\Omega|=2$ can be regarded as a special case when $\mu_t(\omega_1)=0$ for all $t\geq 0$, without loss of generality. Hence $\tilde{\mu}(\omega_1)=\lim\limits_{t\rightarrow\infty}\mu_t(\omega_1)=0$.\\
First, we exclude the case that there exists $\omega_1\in\Omega$ such that $\tilde{\mu}(\omega_1)=0$ and $\tilde{\mu}(\omega)\in(0,1)$ for $\omega\in\{\omega_1,\omega_2\}$:\\
Then by \hyperlink{lem2}{Lemma 2}, there exists $s\in\mathbb{S}$ such that $f(s|\omega_2)>f(s|\omega_3)$. So if $s$ is observed, then
$$\frac{\tilde{\mu}(\omega_2|s)}{\tilde{\mu}(\omega_2)}>\frac{\tilde{\mu}(\omega_3|s)}{\tilde{\mu}(\omega_3)}$$
where $\tilde{\mu}(\omega_2|s)+\tilde{\mu}(\omega_3|s)=\tilde{\mu}(\omega_2)+\tilde{\mu}(\omega_3)=1$. It implies that
$\tilde{\mu}(\omega_2|s)>\tilde{\mu}(\omega_2)$ and $\tilde{\mu}(\omega_3|s)<\tilde{\mu}(\omega_3)$. Then $E_{\tilde{\mu}}[\omega|s]$ will be strictly closer to $\omega_2$ than $E_{\tilde{\mu}}[\omega]$, which violates the  \hyperlink{ntc}{No Trade Condition}.

Then we exclude the case that for all $\omega\in\Omega$, $\tilde{\mu}(\omega)\in(0,1)$.\\
Suppose the case above exists, without loss of generality, suppose $E_{\tilde{\mu}}[\omega]\in[\omega_2,\omega_3)$. So $$(\omega_2-\omega_1)\tilde{\mu}(\omega_1)\leq (\omega_3-\omega_2)\tilde{\mu}(\omega_3)$$
By \hyperlink{lem2}{Lemma 2}, there exists $s_1\in\mathbb{S}$ such that $f(s_1|\omega_1)<f(s_1|\omega_3)$, then
$$\frac{\tilde{\mu}(\omega_3|s_1)}{\tilde{\mu}(\omega_3)}>\frac{\tilde{\mu}(\omega_1|s_1)}{\tilde{\mu}(\omega_1)}$$
Combine the two inequalities above, we have the following:
\begin{align*}
	\tilde{\mu}(\omega_3|s_1)-\tilde{\mu}(\omega_3)&=\Big[	\frac{\tilde{\mu}(\omega_3|s_1)}{\tilde{\mu}(\omega_3)}-1\Big]\tilde{\mu}(\omega_3)\cdot\frac{\tilde{\mu}(\omega_1)}{\tilde{\mu}(\omega_1)}\\
	&>\Big[\frac{\tilde{\mu}(\omega_1|s_1)}{\tilde{\mu}(\omega_1)}-1\Big]\cdot\frac{\tilde{\mu}(\omega_3)}{\tilde{\mu}(\omega_1)}\tilde{\mu}(\omega_1)\\
	&\geq \Big[\frac{\tilde{\mu}(\omega_1|s_1)}{\tilde{\mu}(\omega_1)}-1\Big]\cdot\frac{\omega_2-\omega_1}{\omega_3-\omega_2}\tilde{\mu}(\omega_1)\\
	&=\frac{\omega_2-\omega_1}{\omega_3-\omega_2}\big[\tilde{\mu}(\omega_1|s_1)-\tilde{\mu}(\omega_1)\big]
\end{align*}
Now we check the \hyperlink{ntc}{No Trade Condition}:
\begin{align*}
	E_{\tilde{\mu}}[\omega|s_1]-E_{\tilde{\mu}}[\omega]=&\omega_1\cdot\big[\tilde{\mu}(\omega_1|s_1)-\tilde{\mu}(\omega_1)\big]+\omega_2\cdot\big[\tilde{\mu}(\omega_2|s_1)-\tilde{\mu}(\omega_2)\big]+\omega_3\cdot\big[\tilde{\mu}(\omega_3|s_1)-\tilde{\mu}(\omega_3)\big]\\
	=&(\omega_1-\omega_2)\cdot\big[\tilde{\mu}(\omega_1|s_1)-\tilde{\mu}(\omega_1)\big]+(\omega_3-\omega_2)\cdot\big[\tilde{\mu}(\omega_3|s_1)-\tilde{\mu}(\omega_3)\big]\\
	=&(\omega_3-\omega_2)\cdot\Bigg[\tilde{\mu}(\omega_3|s_1)-\tilde{\mu}(\omega_3)-\frac{\omega_2-\omega_1}{\omega_3-\omega_2}\big[\tilde{\mu}(\omega_1|s_1)-\tilde{\mu}(\omega_1)\big]]\\
	>&0
\end{align*}
So the \hyperlink{ntc}{No Trade Condition} is violated. When $E_{\tilde{\mu}}[\omega]\in(\omega_1,\omega_2)$, we can find a signal $s_2$ such that $f(s_2|\omega_1)>f(s_2|\omega_3)$ and do the same reasoning. Note that in the case where for all $\omega\in\Omega$, $\tilde{\mu}(\omega)\in(0,1)$, $E_{\tilde{\mu}}[\omega]$ cannot be either $\omega_1$ or $\omega_3$.\\
Hence that for all $\omega\in\Omega$, $\tilde{\mu}(\omega)\in\{0,1\}$ is the only possibility.\\
\end{proof}

\noindent \textbf{Proof for \hyperlink{lem5}{Lemma 5}}

\begin{proof}
First we solve the following problem:
$$\max\limits_{\omega}\frac{\tilde{\mu}(\omega)}{\mu_0(\omega)}$$
The solution, $\tilde{\mu}^{-1}(1)$, is unique and the objective function will be $\frac{1}{\mu_0(\tilde{\mu}^{-1}(1))}>0$. And the objective function is just $0$ for all $\omega\in\Omega\backslash\{\tilde{\mu}^{-1}(1)\}$
The objective function can be rewritten as\\
\begin{align*}
	\frac{\tilde{\mu}(\omega)}{\mu_0(\omega)}&=\lim\limits_{t\rightarrow\infty}\frac{\mu_t(\omega)}{\mu_{t-1}(\omega)}\cdot\frac{\mu_{t-1}(\omega)}{\mu_{t-2}(\omega)}\cdots\frac{\mu_1(\omega)}{\mu_{0}(\omega)}\\
	&=\prod_{t=0}^{\infty}\frac{f(S^{a_{t+1}}_{t+1}|\omega)}{\sum_{w=0}^{2}\mu_{t}(w)f(S^{a_{t+1}}_{t+1}|w)}\\
	&=\frac{\prod_{t=0}^{\infty}f(S^{a_{t+1}}_{t+1}|\omega)}{\prod_{t=0}^{\infty}\sum_{w=0}^{2}\mu_{t}(w)f(S^{a_{t+1}}_{t+1}|w)}
\end{align*}
Note that the denominator does not depend on $\omega$, so we just need to maximize $\prod_{t=1}^{\infty}f(S^{a_{t}}_t|\omega)$.
$$\arg\max\limits_{\omega\in\Omega}\prod_{t=1}^{\infty}f(S^{a_{t}}_t|\omega)=\arg\max\limits_{\omega\in\Omega}	\frac{\tilde{\mu}(\omega)}{\mu_0(\omega)}$$

Suppose the true value of the asset is $\omega^*$, then by Jensen's inequality, for all $t>0$ and $\omega\in\Omega$,
$$E_a\Big[\log\frac{f(S^{a_{t}}_t|\omega)}{f(S^{a_{t}}_t|\omega^*)}\Big]\leq\log E_a\Big[\frac{f(S^{a_{t}}_t|\omega)}{f(S^{a_{t}}_t|\omega^*)}\Big]=\log\sum_{a_t\in A}\frac{f(S^{a_{t}}_t|\omega)}{f(S^{a_{t}}_t|\omega^*)}f(S^{a_{t}}_t|\omega^*)=0$$
where $f(S^{a_{t}}_t|\omega^*)$ is the probability that an informative agent take action $a_t$.
By the linearity of expectation and Fatou's Lemma,

\begin{align*}
	E_a\Big[\sum_{t=1}^\infty\log\frac{f(S^{a_{t}}_t|\omega)}{f(S^{a_{t}}_t|\omega^*)}\Big]
	&=E_a\Big[\liminf\limits_{T}\sum_{t=1}^T\log\frac{f(S^{a_{t}}_t|\omega)}{f(S^{a_{t}}_t|\omega^*)}\Big]\\
	&\leq\liminf\limits_{T} E_a\Big[\sum_{t=1}^T\log\frac{f(S^{a_{t}}_t|\omega)}{f(S^{a_{t}}_t|\omega^*)}\Big]\\
	&=\sum_{t=1}^\infty E_a\Big[\log\frac{f(S^{a_{t}}_t|\omega)}{f(S^{a_{t}}_t|\omega^*)}\Big]\\
	&\leq 0
\end{align*}
The equality holds if and only if $\omega=\omega^*$. Hence $\omega^*$ is the solution to maximize
\begin{align*}
	\omega^*&=\arg\max\limits_{\omega\in\Omega}\sum_{t=1}^\infty E_a\Big[\log f(S^{a_{t}}_t|\omega)\Big]\\
	&=\arg\max\limits_{\omega\in\Omega}E_a\Big[\log\prod_{t=1}^{\infty}f(S^{a_{t}}_t|\omega)\Big]\\
\end{align*}
Because $\tilde{\mu}^{-1}(1)=\arg\max\limits_{\omega\in\Omega}	\frac{\tilde{\mu}(\omega)}{\mu_0(\omega)}=\arg\max\limits_{\omega\in\Omega}\prod_{t=1}^{\infty}f(S^{a_{t}}_t|\omega)$, $\tilde{\mu}^{-1}(1)$ should also solve $$\max\limits_{\omega\in\Omega}E_a\Big[\log\prod_{t=1}^{\infty}f(S^{a_{t}}_t|\omega)\Big]$$ Hence $\omega^*=\tilde{\mu}^{-1}(1)\Rightarrow \tilde{\mu}(\omega^*)=1$, the market gets learning!
\end{proof}

\bibliography{references}
\bibliographystyle{apalike}
\end{document}